\documentclass[conference]{IEEEtran}
\IEEEoverridecommandlockouts
\usepackage{cite}
\usepackage{amsmath,amssymb,amsfonts}
\usepackage{algorithmic}
\usepackage{graphicx}
\usepackage{textcomp}
\usepackage{xcolor}
\def\BibTeX{{\rm B\kern-.05em{\sc i\kern-.025em b}\kern-.08em
    T\kern-.1667em\lower.7ex\hbox{E}\kern-.125emX}}
\usepackage{ascmac}
\usepackage{comment}

\usepackage{afterpage}
\usepackage{cite}
\usepackage[utf8]{inputenc}
\usepackage{listings}
\usepackage{multirow}
\usepackage{textcomp}
\usepackage{url}
\usepackage{listings}
\def\RQOne{How does academic literature address gender aspects in software development and engineering in Asia?}
\def\RQTwo{What are the overall trends of software engineering research and practices about gender aspects in Asia?}
\def\RQThree{What are the possible research directions?}
\newcommand{\answer}[2]{~\\[-0.4\baselineskip]\noindent\fbox{\parbox{\columnwidth}{#1 \textbf{#2}}}}

\begin{document}

\title{Systematic Literature Review of Gender and Software Engineering in Asia}

\author{\IEEEauthorblockN{Hironori Washizaki}
\IEEEauthorblockA{Waseda University / National Institute of Informatics / SYSTEM INFORMATION / eXmotion\\
washizaki@waseda.jp}
}

\maketitle

\begin{abstract}
It is essential to discuss the role, difficulties, and opportunities concerning people of different gender in the field of software engineering research, education, and industry. Although some literature reviews address software engineering and gender, it is still unclear how research and practices in Asia exist for handling gender aspects in software development and engineering. We conducted a systematic literature review to grasp the comprehensive view of gender research and practices in Asia. We analyzed the 32 identified papers concerning countries and publication years among 463 publications. Researchers and practitioners from various organizations actively work on gender research and practices in some countries, including China, India, and Turkey. We identified topics and classified them into seven categories varying from personal mental health and team building to organization. Future research directions include investigating the synergy between (regional) gender aspects and cultural concerns and considering possible contributions and dependency among different topics to have a solid foundation for accelerating further research and getting actionable practices. 
\end{abstract}


\IEEEpeerreviewmaketitle

\section{Introduction}
\label{Section: Introduction}

With the social awareness of diversity and inclusion, it is important to discuss the role, difficulties, and opportunities concerning people of different gender in the field of software engineering research, education, and industry \cite{GE2022}. Although some literature reviews address software engineering and gender, such as gender-inclusive requirements engineering \cite{DBLP:conf/er/AlbuquerqueMAGG21}, perceived diversity in software engineering \cite{DBLP:journals/ese/Rodriguez-Perez21}, diversity and agile methodologies \cite{DBLP:conf/icse/SilveiraP19}, and software engineering in digital transformation and diversity \cite{DBLP:conf/compsac/Washizaki22}, 
still it is unclear how research and practices in Asia exist for handling gender aspects in software development and engineering. Since cultures and people in particular regions are different from other regions in many ways that would affect gender aspects, it is worth grasping the big picture of research and practices related to gender aspects reported by researchers and practitioners in Asia. 

We conducted a systematic literature review and answered the following research questions to grasp the comprehensive view of gender research and practices in Asia. 

\begin{itemize}
\item[{\bf RQ1.}] {\bf \RQOne{}} 
To answer this question, we conducted a literature review of the academic literature. We analyzed the 32 identified papers with respect to countries and publication years among 463 publications.

\item[{\bf RQ2.}] {\bf \RQTwo{}} 
To answer this question, we identified topics and classified them into seven categories varying from personal mental health and team building to an organization and gender aspects. 

\item[{\bf RQ3.}] {\bf \RQThree{}}
To answer this question, we summarized the remaining issues and suggested future directions.

\end{itemize}

The rest of this paper is organized as follows. Section II presents the process of our literature review. Section III shows and discusses the review results. Section IV concludes this paper and provides future works. 

\section{Systematic literature review}

We performed a systematic literature review (SLR) to collect research and practices about gender aspects in software development and study reported by researchers and practitioners in Asia. 
An SLR aims to assess scientific papers and group concepts around a topic. 
We chose Scopus\footnote{https://www.scopus.com/} as the search engine since it is effectively used in SLRs of software engineering \cite{DBLP:journals/iotj/WashizakiOHOFY20,DBLP:journals/information/WashizakiXKFKKY21,DBLP:journals/computer/WashizakiKGTNDO22}, and the search results can be exported. 
The database covers many major publishers, including IEEE, ACM, Springer Nature, Wiley Blackwell, Taylor \& Francis, and Elsevier. Furthermore, the database provides a mechanism to perform keyword searches. 

Our process has four steps (1)--(4):

\noindent (1) Initial Search: We executed the following query on titles, abstracts, and keywords of papers regardless of time and subject area. The query specified papers that contain "software engineering" or "software development" from computer science or engineering areas written in English. We used no publication period restrictions. We found 463 publications published from 1994 to 2022. 

\begin{verbatim}
(TITLE-ABS-KEY(gender 
 AND ("software engineering" 
  OR "software development")) 
 AND ( LIMIT-TO ( SUBJAREA,"COMP" ) 
  OR LIMIT-TO ( SUBJAREA,"ENGI" ) ) 
 AND ( LIMIT-TO ( LANGUAGE,"English" ) ) )
\end{verbatim}

\noindent (2) Impurity Removal: Due to the nature of the involved data source, the search results included entities that are not research papers, such as abstracts. We also removed papers written by authors whose affiliations are unspecified or outside Asia. 
Removing such results left 98 papers.  
    
\noindent (3) Inclusion and Exclusion Criteria: For each paper, we vetted whether they should be included in our SLR by applying the following criteria. The titles and abstracts followed by the entire paper were read to determine whether the paper pertained to gender aspects in software development and engineering. Using the definition of our criteria, 32 scholarly papers \cite{DBLP:conf/icse/TahsinAAS22,DBLP:conf/icse/QiuWN21,5458509,DBLP:conf/icis/WangK11,DBLP:conf/sigsoft/WangZ20,10.1007/978-3-642-29458-7_48,DBLP:conf/apsec/KumarSS16,Garg18,DBLP:conf/icse/TiwariASS18,DBLP:journals/tse/PranaFRLPN22,10.1007/978-3-030-68201-9_74,Magableh2019ANEO,DBLP:conf/fie/ZeidE11,DBLP:conf/simultech/SmithTCL11,Mahmod_Md_Dahalin_2012,Gilal19,DBLP:journals/infsof/GilalJOBW16,10.1007/978-981-13-1799-6_38,Gilal16,Gilal19b,DBLP:journals/smr/GilalJCOBA18,Gilal14,Gilal17,7594031,DBLP:conf/chi/MasoodKSIBS21,DBLP:conf/latice/HabibAUR14,DBLP:journals/access/FarooqJMWZA22,DBLP:conf/wcre/SharmaPSNL22,DBLP:journals/imds/LiangLLL07,SAHIN2011916,Fernandez-Sanz12,4717930} were identified.

\begin{itemize}
        \item Inclusion: Papers addressing gender aspects in software development and engineering (particularly developers and teams) research, practices, and education, which are written in English 
        \item Exclusion: Papers focusing on software handling gender information (such as software tools for gender recognition), papers addressing gender biases in software-supported tasks such as resume screening and emotion classification, papers addressing gender differences in pedagogical outcomes of education in general (except for education dedicated to software development), duplicate papers of the same study, or papers that are not written in English
        
\end{itemize}
    
\noindent (4) Data Extraction:  The following information was collected from each paper to answer the research questions: Publication title, publication year, publication venue, author affiliation countries, and topics addressed. 

\section{Results and discussions}

We present and discuss the results of our literature review aligned with the three research questions below. 

\subsection{RQ1. \RQOne{}}

Identified papers by country are shown in Table \ref{Table: List of papers}. As shown in the table, 13 papers were written by authors from Pakistan and Malaysia. However, almost all of them have been written by the same author group, so it does not necessarily imply active gender research over organizations in these countries. 

Other major countries are China, India, and Turkey. Researchers and practitioners from various different organizations are actively working on gender research and practices in these countries. It might reflect that there are much gender and cultural diversity in these countries compared with others. 

\begin{figure}[t]
\vspace{-1.0ex}
\centering
\includegraphics[width=0.9\columnwidth]{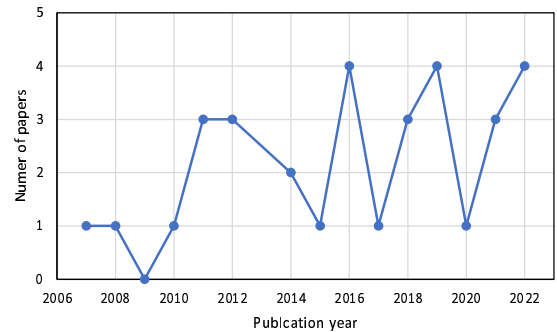}
\vspace{-1.0ex}
\caption{Numbers of papers per year}
\label{Figure: Trend Numbers per Year}
\vspace{-1.0ex}
\end{figure}

Figure \ref{Figure: Trend Numbers per Year} shows the annual trend in the number of papers. Most papers are from 2011 or later, indicating that gender initiatives are emerging in response to the recent increase in social awareness of diversity in the last decade.

\answer{RQ1. \RQOne{}}{
Gender initiatives are emerging in the last decade. 32 academic papers related to gender aspects in software development and engineering were reported from Asia. Researchers and practitioners from various different organizations are actively working on gender research and practices in some countries, including China, India, and Turkey. 
}

\begin{table}[tb]
\begin{center}
\caption{Identified papers on gender aspects in software development and engineering}
\vspace{-1.0ex}
\label{Table: List of papers}
\begin{tabular}{c|cc}
\hline \hline 
Country & Number of papers & References \\ 
\hline
Pakistan & 12 & \cite{Gilal19,DBLP:journals/infsof/GilalJOBW16,10.1007/978-981-13-1799-6_38,Gilal16,Gilal19b,DBLP:journals/smr/GilalJCOBA18,Gilal14,Gilal17,7594031,DBLP:conf/chi/MasoodKSIBS21,DBLP:conf/latice/HabibAUR14,DBLP:journals/access/FarooqJMWZA22}\\
Malaysia & 10 & \cite{Mahmod_Md_Dahalin_2012,Gilal19,DBLP:journals/infsof/GilalJOBW16,10.1007/978-981-13-1799-6_38,Gilal16,Gilal19b,DBLP:journals/smr/GilalJCOBA18,Gilal14,Gilal17,7594031}\\
China & 5 & \cite{DBLP:conf/icse/QiuWN21,5458509,DBLP:conf/icis/WangK11,DBLP:conf/sigsoft/WangZ20,10.1007/978-3-642-29458-7_48}\\
India & 4 & \cite{DBLP:conf/apsec/KumarSS16,Garg18,DBLP:conf/icse/TiwariASS18,DBLP:journals/tse/PranaFRLPN22}\\
Turkey & 3 & \cite{SAHIN2011916,Fernandez-Sanz12,4717930}\\
Kuwait & 2 & \cite{DBLP:conf/fie/ZeidE11,DBLP:conf/simultech/SmithTCL11}\\
Singapore & 2 & \cite{DBLP:journals/tse/PranaFRLPN22,DBLP:conf/wcre/SharmaPSNL22}\\
Bangladesh & 1 & \cite{DBLP:conf/icse/TahsinAAS22}\\
Israel & 1 & \cite{10.1007/978-3-030-68201-9_74}\\
Jordan & 1 & \cite{Magableh2019ANEO}\\
Saudi Arabia & 1 & \cite{DBLP:journals/access/FarooqJMWZA22}\\
South Korea & 1 & \cite{DBLP:journals/access/FarooqJMWZA22}\\
Taiwan & 1 & \cite{DBLP:journals/imds/LiangLLL07}\\
\hline 
\end{tabular}
\end{center}
\vspace{-1.0ex}
\end{table}

\subsection{RQ2. \RQTwo{}}

By carefully reading all identified papers, we identified topics and classified them into the following seven major categories $T_{1}-T_{7}$ varying from personal mental health and team building to organization and gender aspects. 

\begin{itemize}

\item $T_1$. Team building: 14 papers address the team's gender diversity and software engineering-related task performance. Findings and practices can be used for designing better team formation and anticipating team performance  \cite{DBLP:journals/imds/LiangLLL07,5458509,SAHIN2011916,Fernandez-Sanz12,Gilal14,DBLP:conf/fie/ZeidE11,DBLP:journals/infsof/GilalJOBW16,Gilal16,7594031,Gilal17,DBLP:journals/smr/GilalJCOBA18,Gilal19,10.1007/978-981-13-1799-6_38,DBLP:conf/sigsoft/WangZ20}. 
\item $T_2$. Software engineering (SE) education: Six papers address gender differences in software engineering-related education outcomes \cite{4717930,DBLP:conf/icis/WangK11,DBLP:conf/fie/ZeidE11,DBLP:conf/icse/TiwariASS18,10.1007/978-3-030-68201-9_74,DBLP:conf/chi/MasoodKSIBS21}. Two of them are about single-gender classrooms \cite{DBLP:conf/fie/ZeidE11,10.1007/978-3-030-68201-9_74} due to religious and cultural policies. 
\item $T_3$. SE major and job: Three papers discuss the gender and motivation of software engineering-related majors and jobs 
\cite{DBLP:conf/latice/HabibAUR14,DBLP:conf/icse/QiuWN21,DBLP:conf/icse/TahsinAAS22}.
\item $T_4$. OSS and community: Three papers discuss the gender diversity (especially participation of women) in OSS development as well as community support \cite{Mahmod_Md_Dahalin_2012,DBLP:journals/tse/PranaFRLPN22,DBLP:conf/wcre/SharmaPSNL22}. 
\item $T_5$. Mental health: Two papers address gender, mental health, and pressure in software engineering \cite{Garg18,Gilal19b}. 
\item $T_6$. Organization: Two papers handle organizational human resource development and management with gender consideration \cite{10.1007/978-3-642-29458-7_48,DBLP:journals/access/FarooqJMWZA22}. 
\item $T_7$. Others (Research and survey): The paper \cite{Magableh2019ANEO} studied the research-industry collaboration and gender, while another \cite{DBLP:conf/apsec/KumarSS16} surveyed researchers in a specific conference with their demographic attributes including gender. 

\end{itemize}

\answer{RQ2. \RQTwo{}}{
There are seven major topic categories varying from personal mental health and team building to organization and gender aspects. These are team building, SE education, SE major and job, OSS and community, mental health, organization, and others. Although Asian cultures and gender aspects are expected to be related to each other, only few research and practices (i.e., gender segregation in classrooms \cite{DBLP:conf/icis/WangK11,10.1007/978-3-030-68201-9_74}) seem to be directly related to cultural background.
}

\subsection{RQ3. \RQThree{}}

As mentioned above, gender research and practices seem to be active in a limited number of countries in Asia. Thus, researchers and practitioners in Asia are expected to work more actively on gender aspects, share them, and identify what is specific to each region and/or culture and what can be commonly applied \footnote{There might be many papers addressing gender aspects, which are written in domestic languages. We will consider expansion of our survey to include such domestic papers.}. In relation to that, a comparison with other areas outside Asia is also an expected research direction.

Each topic identified above seems to be conducted independently without consideration of possible contributions and dependency among them. For example, mental health and gender ($T_5$) might be supported by other topics such as OSS and community endeavors ($T_4$). Furthermore, gender diversity in organizational human resource development ($T_6$) can be a prerequisite for building teams having proper gender diversity ($T_1$). Making a holistic view by clarifying them should be another research direction, resulting in a solid foundation for accelerating further research and getting actionable practices.

In addition to gender aspects, some of identified papers address various diversity such as personality types \cite{Garg18}, cultural diversity \cite{DBLP:conf/fie/ZeidE11}, and age \cite{DBLP:journals/access/FarooqJMWZA22}. Clarifying relationships among various diversity factors can be an important (meta) research direction. 

\answer{RQ3. \RQThree{}}{
Researchers and practitioners in Asia are expected to work more actively on gender aspects, share them, and identify what is specific to each region and/or culture and what can be commonly applied. Comparison with other areas outside Asia is another expected research direction. Furthermore, it is also expected to clarify contributions and dependencies among different topics to have a solid foundation for further research and actionable practices. Clarifying relationships among various diversity factors can be a significant (meta) research direction.
}

\section{Conclusion and future work}

In this paper, through a systematic literature review, we report on some trends in research and practice on gender aspects in software development and engineering reported by authors in Asia.

We plan to expand the search query and conduct additional reviews in the future. Comparison with other regions outside Asia and investigation of the synergy between gender aspects and cultural concerns are also our future work.

\bibliographystyle{bibliography/IEEEtran}
\bibliography{bibliography/APSEDEI2022}

\end{document}